\documentclass[12pt]{article}
\usepackage{a4wide}
\usepackage{amssymb}
\usepackage{amsmath}
\usepackage{graphicx}
\usepackage{mathdots}
\usepackage{slashed}
\usepackage{verbatim}
\usepackage{fancyvrb}
\usepackage{xcolor}

\usepackage{hyperref}

\usepackage{IEEEtrantools}

\def\makeatletter{\catcode`\@=11}
\makeatletter
\def\mathbox#1{\hbox{$\m@th#1$}}%
\def\math@ccstyles#1#2#3#4#5#6#7{{\leavevmode
      \setbox0\mathbox{#6#7}%
      \setbox2\mathbox{#4#5}%
      \dimen@ #3%
      \baselineskip\z@\lineskiplimit#1\lineskip\z@
      \vbox{\ialign{##\crcr
             \hfil \kern #2\box2 \hfil\crcr
             \noalign{\kern\dimen@}%
             \hfil\box0\hfil\crcr}}}}
\def\mathaccstyles{\math@ccstyles\maxdimen}
\def\maththroughstyles{\math@ccstyles{-\maxdimen}}
\def\unity%
 {\maththroughstyles{.45\ht0}\z@\displaystyle {\mathchar"006C}\displaystyle 1}

\DeclareMathOperator{\SU}{SU}
\DeclareMathOperator{\su}{su}

\DeclareMathOperator{\Sp}{Sp}
\DeclareMathOperator{\SO}{SO}
\DeclareMathOperator{\U}{U}

\newcommand{\bC}{\mathbb{C}}

\begin{document}

\VerbatimFootnotes

\begin{titlepage}

\hfill QMUL-PH-15-07  \\


\vskip 2cm

\begin{center}
{\Large \bfseries
Rigid Supersymmetry from Conformal Supergravity in Five Dimensions
}

\vskip 1.2cm

Alessandro Pini\textsuperscript{$\spadesuit$,$\clubsuit$,1},
Diego Rodriguez-Gomez\textsuperscript{$\spadesuit$,1},
Johannes Schmude\textsuperscript{$\spadesuit$,1}

\bigskip
\bigskip

\begin{tabular}{c}
\textsuperscript{$\spadesuit$}Department of Physics, Universidad de Oviedo, \\
Avda.~Calvo Sotelo 18, 33007, Oviedo, Spain\\
\\
\textsuperscript{$\clubsuit$}Queen Mary University of London, Centre for Research in String Theory,\\
School of Physics, Mile End Road, London, E1 4NS, England\\
\end{tabular}

\vskip 1.5cm

\textbf{Abstract}
\end{center}

\medskip
\noindent
We study the rigid limit of 5d conformal supergravity with minimal supersymmetry on Riemannian manifolds. The necessary and sufficient condition for the existence of a solution is the existence of a conformal Killing vector. Whenever a certain $\SU(2)$ curvature becomes abelian the backgrounds define a transversally holomorphic foliation. Subsequently we turn to the question under which circumstances these backgrounds admit a kinetic Yang-Mills term in the action of a vector multiplet. Here we find that the conformal Killing vector has to be Killing. We supplement the discussion with various appendices.
\bigskip
\vfill

\footnotetext[1]{pinialessandro@uniovi.es,  d.rodriguez.gomez@uniovi.es, schmudejohannes@uniovi.es}

\end{titlepage}

\setcounter{tocdepth}{1}

\tableofcontents

\section{Introduction}

Over the very recent past much effort has been devoted to the study of supersymmetric gauge theories on general spaces. Part of this interest has been triggered by the development of computational methods allowing to exactly compute certain (supersymmetric) observables, such as the supersymmetric partition function (starting with the seminal paper \cite{Pestun:2007rz}), indices or Wilson loops. This program has been very successfully applied to the cases of 4d and 3d gauge theories, and it is only very recently that the 5d case has been considered (\textit{e.g.} \cite{Hosomichi:2012ek,Kallen:2012va,Qiu:2013pta,Kim:2012gu,Qiu:2014cha,Imamura:2012bm}). On the other hand, it has become clear that the dynamics of 5d gauge theories is in fact very interesting, as, contrary to the naive intuition, at least for the case of supersymmetric theories, they can be at fixed points exhibiting rather amusing behavior  as pioneered in \cite{Seiberg:1996bd}. In particular, these theories often show enhanced global symmetries which can be both flavor-like or spacetime-like. The key observation is that vector multiplets in 5d come with an automatically conserved topological current $j\sim \star F\wedge F$ under which instanton particles are electrically charged. These particles provide extra states needed to enhance perturbative symmetries, both flavor or spacetime -- such as what it is expected to happen in the maximally supersymmetric case, where the theory grows one extra dimension and becomes the $(2,\,0)$ 6d theory. In fact, very recently the underlying mechanism for these enhancements has been considered from various points of view \cite{Lambert:2014jna,Tachikawa:2015mha, Zafrir:2015uaa, Rodriguez-Gomez:2015xwa}.

Five-dimensional gauge theories have a dimensionful Yang-Mills coupling constant which is irrelevant in the IR. Hence they are non-renormalizable and thus \textit{a priori} naively uninteresting. However, as raised above, at least for supersymmetric theories the situation is, on the contrary, very interesting as, by appropriately choosing gauge group and matter content, the $g_{YM}$ coulpling which plays the role of UV cut-off can be removed in such a way that one is left with an isolated fixed point theory \cite{Seiberg:1996bd}. From this perspective, it is natural to start with the fixed point theory and think of the standard gauge theory as a deformation whereby one adds a $g_{YM}^{-2}F^2$ term. In fact, the $g_{YM}^{-2}$ can be thought as the VEV for a scalar in a background vector multiplet. Hence, for any gauge theory arising from a UV fixed point\footnote{Note that the theories outside of this class do require a (presumably stringy) UV completion. Hence the class of theories which we are considering is in fact the most generic class of 5d supersymmetric quantum field theories.} we can imagine starting with the conformal theory including a background vector multiplet such that, upon giving a non-zero VEV to the background scalar, it flows to the desired 5d gauge theory. This approach singles out 5d conformally coupled multiplets as the interesting objects to construct.

As described above, on general grounds considering the theory on arbitrary manifolds is very useful, as for example, new techniques allow for exact computation of supersymmetric observables. The first step in this program is of course the construction of the supersymmetric theory on the given (generically curved) space, which is \textit{per se} quite non-trivial. However, the approach put forward by \cite{Festuccia:2011ws} greatly simplifies the task. The key idea is to consider the combined system of the field theory of interest coupled to a suitable supergravity, which, by definition, preserves supersymmetry in curved space. Then, upon taking a suitable rigid limit freezing the gravity dynamics, we can think of the solutions to the gravity sector as providing the background for the dynamical field theory of interest. Note that, since the combined supergravity+field theory is considered off-shell, both sectors can be analyzed as independent blocks in the rigid limit, that is, one can first solve for the supergravity multiplet and then regard such solution as a frozen background for the field theory, where the supergravity background fields act as supersymmetric couplings. Of course, the supergravity theory to use must preserve the symmetries of the field theory which, at the end of the day, we are interested in. Hence, in the case of 5d theories, it is natural to consider conformal supergravity coupled to the conformal matter multiplets described above.

Following this approach, in this paper we will consider 5d conformal supergravity \cite{Fujita:2001kv,Bergshoeff:2001hc,Bergshoeff:2004kh} coupled to 5d conformal matter consisting of both vector and hyper multiplets. As remarked above, the Yang-Mills coupling constant is dimensionful. Hence, the action for the vector multiplets is not the standard quadratic one with a Maxwell kinetic term but rather a cubic action which can be thought as the supersymmetric completion of 5d Chern-Simons. As anticipated, in the rigid limit we can separate the analysis of the gravity multiplet as providing the supersymmetric background for the field theory.
One is thus prompted to study the most generic backgrounds where 5d gauge theories with $\mathcal{N}=1$ supersymmetry can be constructed by analyzing generic solutions of the 5d $\mathcal{N}=2$ conformal supergravity. Solutions to various 5d supergravities on (pseudo-) Riemannian manifolds have been studied in different approaches in \cite{Gauntlett:2002nw,Pan:2013uoa,Imamura:2014ima,Kuzenko:2014eqa,Pan:2014bwa,Alday:2015lta,Pan:2015nba}. For $\mathcal{N}=1$ Poincar\'e supergravity, the necessary and sufficient condition for the existence of a global solution is the existence of a non-vanishing Killing vector. If one considers conformal supergravity this condition becomes the existence of a conformal Killing vector (CKV).\footnote{
  \label{fn:caveat_vanishing_spinor}
  This statement assumes the spinor -- and thus the vector -- to be non-vanishing. In the case of Poincar\'e supergravity, this is always the case if the manifold is connected. After all, the relevant KSE is of the form $\partial_\mu \epsilon^i = \mathcal{O}(\epsilon^i)$. If the spinor vanishes at a point, it vanishes on the whole manifold. For conformal supergravity however, the KSE takes the form of a twistor equation, $\partial_\mu \epsilon^i - \frac{1}{4} \gamma_{\mu\nu} \partial^\nu \epsilon^i = \mathcal{O}(\epsilon^i)$, which has non-trivial solutions even if the right hand side vanishes. The simplest example of this is given by the superconformal supersymmetry in $\mathbb{R}^5$. See section \ref{sec:example_flat_R5}. Here $\epsilon^i \vert_{x^\mu = 0} = v \vert_{x^\mu = 0} = 0$, yet the global solution is non-trivial. In such cases a more careful analysis is necessary.
}
In this paper we analyze Euclidean solutions of 5d conformal supergravity in terms of component fields. Our analysis proceeds along the lines of \cite{Klare:2013dka}.
Interestingly, by studying the conditions under which a VEV for the scalar in the background vector multiplets paying the role of $g_{YM}^{-2}$ can be given in a supersymmetric way, we find that such vector must be in fact Killing. Hence in this case we simply recover the results obtained using Poincar\'e supergravity. 

Our results rely on some reality conditions satisfied by the supersymmetry spinors. In the Lorentzian theory, the spinors generally satisfy a symplectic Majorana condition \eqref{eq:symplectic_Majorana}. If one imposes the same condition in the Euclidean case, there are immediate implication for the spinor bilinears \eqref{eq:spinor_bilinears} that play an important role in the analysis. Namely, the scalar bilinear $s$ is real and vanishes if and only if the spinor vanishes, while the vector bilinear $v$ -- the aforementioned CKV -- is real. One should note however that the symplectic Majorana condition is not equivalent to these conditions for $s$ and $v$. Instead, \eqref{eq:symplectic_Majorana} is slightly stronger, while our results only depend on the milder assumptions on the bilinears.

While the existence of the CKV is a necessary and sufficient condition many of the backgrounds exhibit a more interesting geometric structure -- that of a transversally holomorphic foliation (THF). These appeared already in the context of rigid supersymmetry in three dimensions \cite{Closset:2012ru} and one can think of it as an almost complex structure on the space transverse to the CKV that satisfies a certain integrability condition. A simple example of a five manifold endowed with a THF is given by Sasakian manifolds. Here, the existence of the THF was exploited in \cite{Schmude:2014lfa} in order to show that the perturbative partition function can be calculated by counting holomorphic functions on the associated K\"ahler cone. Similar considerations were used in \cite{Pan:2014bwa} to solve the BPS equations on the Higgs branch. This gives rise to the question whether such simplifications occur in localization calculations on more generic five manifolds admitting rigid supersymmetry. This was addressed in \cite{Pan:2015nba} in the context of 5d $\mathcal{N}=1$ Poincar\'e supergravity. Here it was shown that a necessary and sufficient condition for such manifolds to admit a supersymmetric background is the existence of a Killing vector. If an $\su(2)$-valued scalar in the Weyl multiplet is non-vanishing and covarinatly constant along the four-dimensional leaves of the foliation it follows furthermore that the solution defines a THF. Subsequently it was argued that the existence of a THF (or that of an integrable Cauchy-Riemann structure) is sufficient to lead to similar simplifications in the context of localization as in \cite{Pan:2014bwa,Schmude:2014lfa}.

With this motivation in mind we will address the question under which circumstances generic backgrounds of the conformal supergravity in question admit THFs. Our results are to be seen in the context of the very recent paper \cite{Alday:2015lta}. We will find that the necessary and sufficient condition for the solution to support a THF is the existence of a global section of an $\su(2)/\mathbb{R}$ bundle that is covariantly constant with respect to a connection $\mathcal{D}^Q$ that arises from the intrinsic torsions parametrizing the spinor.

The outline of the rest of the paper is as follows. In section \ref{sec: SUGRA_review} we offer a lightning review of the relevant aspects of  superconformal 5d supergravity, with our conventions compiled in appendix \ref{sec:conventions} and further details described in appendix \ref{sec:computation}. In section \ref{sec:general_solutions} we turn to the analysis of the general solutions of the supergravity, showing that the necessary condition for supersymmetry is the presence of a conformal Killing vector. Moreover, we will see that given a Killing spinor and the related CKV the general solution depends only on an $\su(2)$-valued $\Delta^{ij}$ and a vector $W$ that is orthogonal to the CKV. Both are determined by solving simple ODEs that become trivial if one goes to a frame where the CKV is Killing. In section \ref{sec:gauge_theory} we study under which conditions it is possible to turn on a VEV for scalars in background vector multiplets thus flowing to a standard gauge theory, finding that the requirement is that the vector is not only conformal Killing but actually Killing. In section \ref{sec:THF} we derive the conditions for the existence of a THF. In section \ref{sec:examples} we show how some particular examples fit into our general structure, describing in particular the cases of $\mathbb{R}\times S^4$ relevant for the index computation of \cite{Kim:2012gu} and the $S^5$ relevant for the partition function computation of  \cite{Hosomichi:2012ek,Kallen:2012va}. We finish with some conclusions in section \ref{sec:conclusions}.
 
\vspace{1cm}
\textbf{Note added:} While this work was in its final stages we received \cite{Alday:2015lta}, which has a substantial overlap with our results.

\section{Five-dimensional conformal supergravity}\label{sec: SUGRA_review}

Let us begin by reviewing the five-dimensional, $\mathcal{N}=2$ conformal supergravity of \cite{Bergshoeff:2001hc,Bergshoeff:2004kh}\footnote{A word on notation is in order here. We stress that we are discussing minimal supersymmetry in five dimensions.}. The theory has $SU(2)_R$ $R$-symmetry. The Weyl multiplet contains the vielbein $e_\mu^a$, the $\SU(2)_R$ connection $V_\mu^{(ij)}$, an antisymmetric tensor $T_{\mu\nu}$, a scalar $D$, the gravitino $\psi_\mu^i$ and the dilatino $\chi^i$. Our conventions are summarised in appendix \ref{sec:conventions}. 

The supersymmetry variations of the gravitino and dilatino are
\begin{IEEEeqnarray}{rCl}
  \delta \psi_\mu^i &=& \mathcal{D}_\mu \epsilon^i + \imath \gamma \cdot T \gamma_\mu \epsilon^i - \imath \gamma_\mu \eta^i, \label{eq:gravitino-variation}\\
  \delta \chi^i &=& \frac{1}{4} \epsilon^i D - \frac{1}{64} \gamma \cdot \hat{R}^{ij}(V) \epsilon_j + \frac{\imath}{8} \gamma^{\mu\nu} \slashed{\nabla} T_{\mu\nu} \epsilon^i - \frac{\imath}{8} \gamma^\mu \nabla^\nu T_{\mu\nu} \epsilon^i - \frac{1}{4} \gamma^{\kappa\lambda\mu\nu} T_{\kappa\lambda} T_{\mu\nu} \epsilon^i \nonumber\\
  &&+ \frac{1}{6} T^2 \epsilon^i + \frac{1}{4} \gamma \cdot T \eta^i. \label{eq:dilatino-variation}
\end{IEEEeqnarray}
Up to terms $\mathcal{O}(\psi_\mu, \chi^i)$,
\begin{IEEEeqnarray}{rCl}
  \mathcal{D}_\mu \epsilon^i &=& \partial_\mu \epsilon^i + \frac{1}{4} \omega_\mu^{ab} \gamma_{ab} \epsilon^i + \frac{1}{2} b_\mu \epsilon^i - V_\mu^{ij} \epsilon_j, \\
  \hat{R}_{\mu\nu}^{ij}(V) &=& dV^{ij}_{\mu\nu} - 2 V_{[\mu}^{k(i} V_{\nu] k}^{\phantom{\nu] k}j)}.
\end{IEEEeqnarray}
In what follows we will set the Dilation gauge field $b_\mu$ to zero.

As usual, taking the $\gamma$-trace of the gravitino equation allows to solve for the superconformal parameters as $\eta^i = - \frac{\imath}{5} \slashed{\mathcal{D}} \epsilon^i + \frac{1}{5} T \cdot \gamma \epsilon^i$. Hence, we can rewrite the equations arising from the gravitino and dilatino as
\begin{IEEEeqnarray}{rCl}
  0 &=& \mathcal{D}_\mu \epsilon^i - \frac{1}{4} \gamma_{\mu\nu} \mathcal{D}^\nu \epsilon^i + \imath \gamma_{\mu\kappa\lambda} T^{\kappa\lambda} \epsilon^i - 3 \imath T_{\mu\nu} \gamma^\nu \epsilon^i, \label{eq:gravitino-variation_w/out_eta}\\
  0 &=& \frac{1}{128} \epsilon^i (32 D + R) + \frac{1}{15} T_{\mu\nu} T^{\mu\nu} \epsilon^i + \frac{1}{8} \mathcal{D}^\mu \mathcal{D}_\mu \epsilon^i + \frac{3\imath}{40} \gamma_{\kappa\lambda\mu} T^{\kappa\lambda} \mathcal{D}^\mu \epsilon^i + \frac{11\imath}{40} \gamma^\mu T_{\mu\nu} \mathcal{D}^\nu \epsilon^i \nonumber\\
  &&+ \frac{\imath}{4} \gamma_{\mu\kappa\lambda} \nabla^\mu T^{\kappa\lambda} \epsilon^i + \frac{\imath}{2} \gamma^\mu \nabla^\nu T_{\mu\nu} \epsilon^i - \frac{1}{5} \gamma^{\kappa\lambda\mu\nu} T_{\kappa\lambda} T_{\mu\nu} \epsilon^i.\label{eq:dilatino-variation_D-squared-form}
\end{IEEEeqnarray}
Here, $R$ is the Ricci scalar and the rewriting of the dilatino equation uses the gravitino equation. One could also rewrite the latter using $\slashed{\mathcal{D}}^2$ as in \cite{Klare:2013dka}, yet we found the above formulation to be more economical in this case.

\section{General solutions of $\mathcal{N}=2$ conformal supergravity}\label{sec:general_solutions}

General solutions to five-dimensional conformal supergravity have been constructed in \cite{Kuzenko:2014eqa} using superspace techniques. In this section we will provide an alternative derivation of the most general solutions to $\mathcal{N}=2$ conformal supergravity in euclidean signature using component field considerations along the lines of \cite{Klare:2013dka}. Before turning to the details, let us recall a counting argument from \cite{Klare:2013dka} regarding these solutions: In general the gravitino yields $40$ scalar equations. Eliminating the superconformal spinor $\eta^i$ removes $8$. As we will see, the gravitino equation then also fixes the $10$ components of the antisymmetric tensor and $8$ of the components of the $\SU(2)$ connection. This leaves us with $14$, which is exactly enough to remove the traceless, symmetric part of a two-tensor $P$ which will appear in the intrinsic torsion. Since the trace is undetermined we will find a CKV; the vector is Killing if the trace vanishes. The remaining $7$ components of the $\SU(2)$ connection and the scalar in the Weyl multiplet will then be determined by the eight equations arising from the dilatino variation.

In order to study solutions of \eqref{eq:gravitino-variation_w/out_eta} and \eqref{eq:dilatino-variation_D-squared-form}, we introduce the bispinors
\begin{equation}\label{eq:spinor_bilinears}
  \begin{aligned}
    s &= \epsilon^i C \epsilon_{i}, \\
    v &= (\epsilon^i C \gamma_\mu \epsilon_{i}) dx^\mu, \\
    \Theta^{ij} &= (\epsilon^i C \gamma_{\mu\nu} \epsilon^j) dx^\mu \otimes dx^\nu,
  \end{aligned}
\end{equation}
In what follows, we will assume the scalar $s$ to be non-zero and the one-form $v$ to be real. These assumptions are implied if one imposes a symplectic Majorana condition such as \eqref{eq:symplectic_Majorana}. Furthermore, note that $v^2 = s^2$. 

The one-form then decomposes the tangent bundle into a horizontal and a vertical part, with the former being defined as $TM_H = \{ X \in TM \vert v(X) = 0\}$ and $TM_V$ as its orthogonal complement. Due to the existence of a metric we use $v$ to refer both to the one-form and the correspoding vector and an analogous decomposition into horizontal and vertical forms extends to the entire exterior algebra. In turn, the two-forms $\Theta^{ij}$ are  fully horizontal and anti self-dual\footnote{\label{fn:Theta_identities}
Explicitly, the self-duality condition is
\begin{equation*}
  \Theta^{ij}_{\mu\nu} = - \frac{1}{2} s^{-1} \epsilon_{\mu\nu\kappa\lambda\rho} \Theta^{ij\kappa\lambda} v^\rho, \qquad
  \epsilon_{\lambda\mu\nu\sigma\tau} \Theta^{ij\sigma\tau} = -3! s^{-1} \Theta_{[\lambda\mu}^{ij} v_{\nu]}.
\end{equation*}
}
with respect to the automorphism $\iota_{s^{-1} v} \star : \Lambda^2_H \to \Lambda^2_H$:
\begin{equation}
  \iota_{v} \Theta^{ij} = 0, \qquad
  \iota_{s^{-1}v} \star \Theta^{ij} = - \Theta^{ij}.
\end{equation}
One finds that the spinor is chiral with respect to the vector $s^{-1} v$,
\begin{equation}
  s^{-1} v^\mu \gamma_\mu \epsilon^i = \epsilon^i.
\end{equation}
Note that the sign here is mainly a question of convention. Had we defined $v$ with an additional minus sign, we would find the spinor to be anti-chiral and $\Theta^{ij}$ to be self-dual. One can see this by considering the transformation $v \mapsto -v$. In addition, we define the operator $\Pi^\mu_\nu = \delta^\mu_\nu - s^{-2} v^\mu v_\nu$ which projects onto the horizontal space. A number of additional useful identities involving $\Theta^{ij}$ are given in appendix \ref{sec:conventions}.

Next, we parametrize the covariant derivative of the supersymmetry spinor using intrinsic torsions as in \cite{Imamura:2014ima},
\begin{equation}\label{eq:intrinsic_torsions_introduced}
  \nabla_\mu \epsilon^i \equiv P_{\mu\nu} \gamma^\nu \epsilon^i + Q_\mu^{ij} \epsilon_j.
\end{equation}
Here, $P_{\mu\nu}$ is a two-tensor while $Q_\mu^{ij}$ is symmetric in its $SU(2)_R$ indices. Rewriting the torsions in terms of the supersymmetry spinor one finds
\begin{equation}\label{eq:intrinsic_torsions_in_terms_of_covariant_derivative}
  s P_{\mu\nu} = \epsilon^i \gamma_\nu \nabla_\mu \epsilon_i = \frac{1}{2} \nabla_\mu v_\nu, \qquad
  s Q_\mu^{ij} = 2 \epsilon^{(i} \nabla_\mu \epsilon^{j)}.
\end{equation}

\subsection{The gravitino equation}

We now turn to the study of generic solutions of \eqref{eq:gravitino-variation_w/out_eta} and \eqref{eq:dilatino-variation_D-squared-form} using the intrinsic torsions. The reader interested in intermediate results and some technical details might want to consult appendix \ref{sec:computation}. To begin, substituting \eqref{eq:intrinsic_torsions_introduced} and contracting with $\epsilon_i \gamma_\kappa$ as well as $\epsilon^j$ and symmetrizing in $i, j$ one finds that \eqref{eq:gravitino-variation_w/out_eta} is equivalent to
\begin{IEEEeqnarray}{rCl}
  0 &=& \frac{5}{4} s \left( P_{(\mu\nu)} - \frac{1}{5} g_{\mu\nu} P^\lambda_{\phantom{\lambda}\lambda} \right) + \frac{3}{4} s \left( P_{[\mu\nu]} - 4 \imath T_{\mu\nu} \right) + \frac{1}{4} \epsilon_{\mu\nu\kappa\lambda\rho} (P^{[\kappa\lambda]} - 4 \imath T^{\kappa\lambda}) v^\rho \nonumber\\
  &&+ \frac{1}{8} \epsilon_{\mu\nu\rho\sigma\tau} (Q - V)^{\rho ij} \Theta^{\sigma\tau}_{ij}, \label{eq:gravitino_via_torsion_vector_contraction} \\
  0 &=& \frac{1}{2} s (Q - V)_\mu^{ij} + \frac{1}{4} (Q - V)^{\nu (j}_{\phantom{\nu (j}k} \Theta^{i) k}_{\mu\nu} + \frac{1}{8} \epsilon_{\mu\kappa\lambda\sigma\tau} (P - 4 \imath T)^{\kappa\lambda} \Theta^{ij\sigma\tau}. \label{eq:gravitino_via_torsion_triplet_contraction}
\end{IEEEeqnarray}
Clearly, the symmetric part in \eqref{eq:gravitino_via_torsion_vector_contraction} has to vanish independently; so we find
\begin{equation}\label{eq:symmetric_part_of_P}
  P_{(\mu\nu)} = \frac{1}{5} g_{\mu\nu} P^\lambda_{\phantom{\lambda}\lambda}.
\end{equation}
This implies that $v$ is a conformal Killing vector as can be seen using \eqref{eq:intrinsic_torsions_in_terms_of_covariant_derivative}.

By contracting the two remaining equations with $v^\mu$, one finds
\begin{IEEEeqnarray}{rCl}
  0 &=& 3 s v^\mu (P - 4 \imath T)_{[\mu\nu]} - s \Theta^{ij}_{\nu\mu} (Q - V)^\mu_{ij}, \label{eq:QVPT_identity_vector}\\
  0 &=& 2 s v^\mu (Q - V)_\mu^{ij} - s \Theta^{ij}_{\mu\nu} (P - 4 \imath T)^{\mu\nu} \label{eq:QVPT_identity_triplet}.
\end{IEEEeqnarray}
Projecting \eqref{eq:gravitino_via_torsion_vector_contraction} on the horizontal space, we find that $\Pi(P - 4\imath T)$ is anti self-dual.
\begin{equation}
  0 = (P - 4 \imath T)^+.
\end{equation}
Contracting \eqref{eq:QVPT_identity_triplet} with $\Theta_{ij\kappa\lambda}$ and using \eqref{eq:double_theta_identity} gives us the horizontal, self-dual part.
\begin{equation}
  (P - 4\imath T)^- = s^{-2} \Theta^{ij} \imath_v (Q - V)_{ij}.
\end{equation}
By now we have equations for the self-dual, anti self-dual and vertical components of $(P-4\imath T)_{[\mu\nu]}$, which means that all components of this two-form are determined. Putting everything together, we find
\begin{IEEEeqnarray}{rCl}\label{eq:gravitino_solution_i}
  s^2 (P - 4\imath T)_{[\mu\nu]} &=& \frac{1}{3} \left[ (v \wedge \Theta^{ij})_{\mu\nu\rho} + 2 \Theta_{\mu\nu}^{ij} v_\rho\right] (Q-V)^\rho_{ij}.
\end{IEEEeqnarray}

The only equation we have not considered so far is the horizontal projection of \eqref{eq:gravitino_via_torsion_triplet_contraction}. After using \eqref{eq:double_theta_identity_2}, \eqref{eq:QVPT_identity_vector} and \eqref{eq:QVPT_identity_triplet} this simplifies to
\begin{equation}\label{eq:gravitino_solution_ii}
  s \Pi_\mu^{\phantom{\mu}\nu} (Q-V)_{\nu\phantom{i}j}^{\phantom{\nu}i} = -\frac{1}{2} [(Q-V)^\nu, \Theta_{\mu\nu}]^i_{\phantom{i}j}.
\end{equation}
In summary, the gravitino is solved by \eqref{eq:gravitino_solution_i} and \eqref{eq:gravitino_solution_ii}.

Note that one can solve \eqref{eq:gravitino_solution_ii} by brute force after picking explicit Dirac matrices. One finds that the equation leaves seven components of $(Q-V)$ unconstrained. Three of these have to be parallel to $v$ as they do not enter in \eqref{eq:gravitino_solution_ii}. This suggests that it is possible to package the seven missing components into a triplet $\Delta^{ij}$ (three components) and a horizontal vector $W^\mu$ (four) and parametrize a generic solution of the gravitino equation as
\begin{equation}\label{eq:Q-V_parametrization}
  (Q - V)_\mu^{ij} = s^{-1} \left( v_\mu \Delta^{ij} + W^\lambda \Theta^{ij}_{\lambda\mu} \right)
  \qquad \text{s.t.} \qquad
  v(W) = 0, \Delta^{ij} = \Delta^{ji}.
\end{equation}
Using \eqref{eq:double_theta_identity_2} one can verify that \eqref{eq:Q-V_parametrization} satisfies \eqref{eq:gravitino_solution_ii}. The above implies that
\begin{equation}\label{eq:T-solution}
  T_{\mu\nu} = \frac{\imath}{4} \left( s^{-1} \Theta^{ij}_{\mu\nu} \Delta^{ij} + s^{-1} v_{[\mu} W_{\nu]} - P_{[\mu\nu]} \right).
\end{equation}

\subsection{The dilatino equation}\label{sec:dilatino_analysis}

We finally turn to the dilatino equation \eqref{eq:dilatino-variation_D-squared-form}. To begin, we note that between $\Delta^{ij}, W_\mu, D$ there are eight unconstrained functions remaining while the dilatino equation provides eight constraints. We can thus expect that there will be no further constraints on the geometry. In this respect, similarly to \cite{Klare:2013dka}, supersymmetry is preserved as long as the manifold supports a conformal Killing vector $v$. 

In what follows we will need to deal with terms involving derivatives of the spinor bilinears \eqref{eq:spinor_bilinears}. To do so we use the identities
\begin{IEEEeqnarray}{rCl}
  \nabla_\mu s &=& 2 P_{\mu\nu} v^\nu, \\
  \nabla_\mu v_\nu &=& 2 s P_{\mu\nu}, \\
  \nabla_\mu \Theta_{\kappa\lambda}^{ij} &=& 3! s^{-1} \Theta^{ij}_{[\kappa\lambda} v_{\rho]} P_\mu^{\phantom{\mu}\rho} - 2 \Theta_{\kappa\lambda}^{k(i} Q^{j)}_{\mu k}, \\
  \nabla_{[\lambda} P_{\mu\nu]} &=& - s^{-1} P_{[\mu\nu} \nabla_{\lambda]} s.
\end{IEEEeqnarray}

Contracting \eqref{eq:dilatino-variation_D-squared-form} with $\epsilon^j$ and symmetrizing over the $SU(2)_R$ indices $i, j$ one finds
\begin{IEEEeqnarray}{rCl}
  0 &=& \frac{1}{8} \epsilon^{(i} \mathcal{D}^\mu \mathcal{D}_\mu \epsilon^{j)} + \frac{3\imath}{40} \epsilon^{(i} \gamma_{\kappa\lambda\mu} T^{\kappa\lambda} \mathcal{D}^\mu \epsilon^{j)} + \frac{11\imath}{40} \epsilon^{(i} \gamma^\mu T_{\mu\nu} \mathcal{D}^\nu \epsilon^{j)} \nonumber\\
  &&+ \frac{\imath}{4} \epsilon^{(i} \gamma_{\kappa\lambda\mu} \epsilon^{j)} \nabla^\mu T^{\kappa\lambda}.
\end{IEEEeqnarray}
Substituting \eqref{eq:Q-V_parametrization} and \eqref{eq:T-solution} one finds after a lengthy calculation\footnote{We found the Mathematica package \texttt{xAct} \cite{martin2007invar,MartinGarcia:2008qz} very useful.}
\begin{IEEEeqnarray}{rCl}\label{eq:dilatino_solution_Delta_ODE}
  \pounds_v \Delta^i_{\phantom{i}j} &=& -\frac{2}{5} s P^\mu_{\phantom{\mu}\mu} \Delta^i_{\phantom{i}j} - [\iota_v Q + P^{[\mu\nu]}\Theta_{\mu\nu}, \Delta]^i_{\phantom{i}j}.
\end{IEEEeqnarray}

Contracting \eqref{eq:dilatino-variation_D-squared-form} with $-\epsilon_i \gamma_\mu$ one obtains
\begin{IEEEeqnarray}{rCl}\label{eq:dilatino_vector_equation}
  0 &=& v_\mu \left( \frac{32 D + R}{128} + \frac{1}{15} T_{\mu\nu} T^{\mu\nu} \right) + \frac{1}{8} \epsilon^i \gamma_\mu \mathcal{D}^\nu \mathcal{D}_\nu \epsilon_i + \frac{3\imath}{40} \epsilon^i \gamma_\mu \gamma_{\kappa\lambda\nu} T^{\kappa\lambda} \mathcal{D}^\nu \epsilon_i \nonumber\\
  &&+ \frac{11\imath}{40} \epsilon^i \gamma_\mu \gamma^\kappa T_{\kappa\lambda} \mathcal{D}^\lambda \epsilon_i + \frac{\imath}{4} \epsilon_\mu^{\phantom{\mu}\nu\kappa\lambda\sigma} v_\sigma \nabla_\nu T_{\kappa\lambda} + \frac{\imath s}{2} \nabla^\nu T_{\mu\nu} - \frac{s}{5} \epsilon_\mu^{\phantom{\mu}\kappa\lambda\sigma\tau} T_{\kappa\lambda} T_{\sigma\tau}.
\end{IEEEeqnarray}
The vertical component of this fixes the scalar $D$.
\begin{IEEEeqnarray}{rCl}
\label{eq:D}
  0 &=& 480 s D + 15 s R
  + 48 s (P^\mu_{\phantom{\mu}\mu})^2
  - 130 s W^2
  + 60 \epsilon_{\kappa\lambda\mu\nu\rho} P^{[\kappa\lambda]} P^{[\mu\nu]} v^\rho
  - 160 s \Delta^{ij} \Delta_{ij} \nonumber\\
  &&+ 100 P_{[\mu\nu]} (s P^{[\mu\nu]} - 2 v^\mu W^\nu)
  - 200 P^{[\mu\nu]} \Theta_{\mu\nu}^{ij} \Delta_{ij}
  + 48 v^\mu \nabla_\mu P^\rho_{\phantom{\rho}\rho} - 120 s \nabla^\mu W_\mu.
\end{IEEEeqnarray}
The horizontal part of \eqref{eq:dilatino_vector_equation} yields a differential equation for $W$
\begin{IEEEeqnarray}{rCl}
\label{eq:W}
  \pounds_v W_\kappa = \frac{1}{50} \Pi_\kappa^{\phantom{\kappa}\lambda} &(& 3 s^2 P^\mu_{\phantom{\mu}\mu} W_\lambda  - 34 P^\rho_{\phantom{\rho}\rho} P_{[\lambda\mu]} v^\mu 
    - 20 s \nabla_\lambda P^\rho_{\phantom{\rho}\rho} ).
\end{IEEEeqnarray}
Note that the left hand side is horizontal since $\iota_v \pounds_v W = \iota_v \iota_v dW = 0$.

Similar to the discussion in \cite{Klare:2013dka}, we note that one can always solve \eqref{eq:dilatino_solution_Delta_ODE} and \eqref{eq:W} locally. Moreover, after a Weyl transformation to a frame where $v$ is not only conformal Killing yet actually Killing, that is, setting $P^\mu_{\phantom{\mu}\mu}=0$, both equations simplify considerably. All the source terms in the latter vanish which is now solved by $W = 0$ while the former becomes purely algebraic,
\begin{IEEEeqnarray}{rCl}\label{eq:dilatino_solution_Delta_ALGEBRAIC}
0 &=&  [\iota_v Q + P^{[\mu\nu]}\Theta_{\mu\nu}, \Delta]^i_{\phantom{i}j},
\end{IEEEeqnarray}
and is solved by $\Delta = s^{-1} f (\iota_v Q + P^{[\mu\nu]} \Theta_{\mu\nu})$ for a generic, possibly vanishing, function $f$ as long as $\pounds_v f = 0$. The factor $s^{-1}$ is simply included here to render $\Delta$ invariant under $\epsilon^i \to \lambda \epsilon^i$ for $\lambda \in \bC$.

An alternative way to see that \eqref{eq:dilatino_solution_Delta_ODE} and \eqref{eq:W} can be solved globally is by direct construction of the solution following the approach of section 5 in \cite{Pan:2015nba}. Thus, the existence of a non-vanishing CKV is not only necessary, but also sufficient. See also footnote \ref{fn:caveat_vanishing_spinor}.

\section{Yang-Mills theories from conformal supergravity}\label{sec:gauge_theory}

The solutions described above provide the most general backgrounds admitting a five-dimensional, minimally supersymmetric quantum field theory arising in the rigid limit of conformal supergravity. In the maximally supersymmetric case a more general class of solutions is possible, since the R-symmetry of maximal supergravity is $\SO(5)$, one can define supersymmetric field theories on generic five manifolds by twisting with the whole $\SO(5)$. Such field theories were considered in \cite{Bak:2015hba}. An embedding in supergravity should be possible starting from \cite{Cordova:2013bea}.

Of course, in the case at hand our starting point is conformal supergravity, so only conformal multiplets can be consistently coupled to the theory. While the hypermultiplet is conformally invariant \textit{per se}, the vector multiplet with the standard Maxwell kinetic term breaks conformal invariance as the Yang-Mills coupling has negative mass dimensions. Therefore the action for the conformally coupled vector multiplet is a non-standard cubic action which can be thought as the supersymmetric completion of 5d Chern-Simons. Such action contains in particular a coupling of the form $C_{IJK}\,\sigma^I\, F^J\,F^K$, where $F^I$ is the field strength of the $I$-th vector multiplet, $\sigma^I$ its corresponding real scalar and $C_{IJK}$ a suitable matrix encoding the couplings among all vector multiplets (we refer to \cite{Bergshoeff:2001hc,Bergshoeff:2004kh} for further explanations). Thus we can imagine constructing a standard gauge theory by starting with a conformal theory and giving suitable VEVs to scalars in background abelian vector multiplets. Of course, such VEVs must preserve supersymmetry. To that end, let us consider the SUSY variation of a background vector multiplet. As usual, only the gaugino variation is relevant, which, in the conventions of \cite{Bergshoeff:2004kh}, reads
\begin{equation}
\delta\Omega_B^i=-\frac{\imath}{2}\,\slashed{\nabla}\sigma_B\,\epsilon^i+Y_B^i\,_j\,\epsilon^j+\sigma_B\,\gamma\cdot T\epsilon^i+\sigma_B\,\eta^i\, ,
\end{equation}
where we have set to zero the background gauge field. The $Y_B^i\,_j$ are a triplet of auxiliary scalars in the vector multiplet. Contracting with $\epsilon_i$ it is straightforward to see that, in order to have a supersymmetric VEV, we must have
\begin{equation}
\label{constraint}
 \pounds_v \sigma_B+\frac{2\,s}{5}\,P^\mu_{\phantom{\mu}\mu} \sigma_B =0\, ,
\end{equation}
while the other contractions fix the value of $Y_B^i\,_j$. The VEV of $\sigma_B$ is $g_{YM}^{-2}$, and as such one would like it to be a constant. Therefore, equation \eqref{constraint} gives us an obstruction for the existence of a Maxwell kinetic term; namely, that $v$ is Killing and not only conformal Killing. It then follows that all backgrounds admitting standard -- \textit{i.e.} quadratic -- supersymmetric Yang-Mills theories, involve a $v$ which is a genuine Killing vector. They are thus solutions of the $\mathcal{N}=1$ Poincar\'e supergravity -- see \textit{e.g.} \cite{Pan:2013uoa,Imamura:2014ima,Pan:2014bwa,Pan:2015nba}. In particular, the case of $\mathbb{R}\times S^4$ is of special interest as the partition function on this space in the absence of additional background fields gives the superconformal index \cite{Kim:2012gu}. The relevant supersymmetry spinors appearing in the calculation define a vector $v$ which is conformal Killing; and therefore the background is only a solution of conformal supergravity. As we will explicitly see below, it is easy to check that such a solution, which can be easily obtained by a simple change of coordinates in the spinors in \cite{Rodriguez-Gomez:2015xwa}, nicely fits in our general discussion above.  If, on the other hand, one studies supersymmetric backgrounds on $S^5$ without additional background fields, one finds $v$ to be Killing (see below as well). Thus such backgrounds can be regarded as a solution to conformal supergravity that are not obstructed by \eqref{constraint} and do thus admit a constant $\sigma_B$. In fact, it is easy to check this nicely reproduces the results of \cite{Hosomichi:2012ek}.

Eq.~\eqref{constraint} shows that backgrounds admitting only a conformal Killing vector cannot support a standard gauge theory with a constant Maxwell kinetic term. As anticipated above, and explicitly described below, this is precisely the case of $\mathbb{R}\times S^4$, relevant for the computation of the index. Of course it is possible to solve \eqref{constraint} if one accepts that the Yang-Mills coupling is now position dependent. This way we can still think of the standard Yang-Mills action as a regulator to the index computation.\footnote{One might wonder that the cubic lagrangian theory is enough. However, in some cases such as \textit{e.g.} $\Sp$ gauge theories, such cubic lagrangian is identically zero.} While this goes beyond the scope of this paper, one might imagine starting with the Yang-Mills theory on $\mathbb{R}^5$ where \eqref{constraint} can be satisfied for a constant $\sigma_B$. Upon conformally mapping $\mathbb{R}^5$ into $\mathbb{R}\times S^4$ the otherwise constant $\sigma_B=g_{YM}^{-2}$ becomes $\sigma_B=g_{YM}^{-2}\,e^{\tau}$, being $\tau$ the coordinate parametrizing $\mathbb{R}$. In the limit $g_{YM}^{-2}\rightarrow 0$ we recover the conformal theory of \cite{Kim:2012gu}. One can imagine computing the supersymmetric partition function in this background. As the preserved spinors are just the same as in the $g_{YM}^{-2}\rightarrow 0$ limit, the localization action, localization locus and one-loop fluctuations will be just the same as in the conformal case. While we leave the computation of the classical action for future work, it is clear that the limit $g_{YM}^{-2}\rightarrow 0$ will reproduce the result in  \cite{Kim:2012gu}.

\section{Existence of transversally holomorphic foliations}\label{sec:THF}

We will now discuss under which circumstances solutions to equations \eqref{eq:gravitino-variation} and \eqref{eq:dilatino-variation} define transversally holomorphic foliations (THF). Since we assumed $s \neq 0$ and $v$ real, it follows that the CKV $v$ is non-vanishing and thus that $v$ defines a foliation on $M$. Using \eqref{eq:double_theta_identity_2} one can show then that $\Theta^{ij}$ defines a triplet of almost complex structures on the four-dimensional horizontal space $TM_H$. Thus, given a non-vanishing section\footnote{
  Since $\Phi$ is invariant under $m_{ij} \mapsto f m_{ij}$ for any non-vanishing function $f:M \to \mathbb{R}$ it might be more appropriate to think of $m_{ij}$ as a ray in the three-dimensional $\su(2)$ vector space. From this point of view, $m_{ij}$ is a map
  \begin{equation*}
    m : M \to S^2 \subset \su(2).
  \end{equation*}
}
of the $\su(2)_R$ Lie algebra $m_{ij}$ we can define an endomorphism on $TM$
\begin{equation}
  (\Phi[m])^\mu_{\phantom{\mu}\nu} \equiv (\det m)^{-1/2} m_{ij} (\Theta^{ij})^\mu_{\phantom{\mu}\nu}.
\end{equation}
This satisfies $\Phi[m]^2 = -\Pi$ and thus induces a decomposition of the complexified tangent bundle
\begin{equation}
   T_{\bC}M = T^{1,0} \oplus T^{0,1} \oplus \bC v.
\end{equation}
Any such decomposition is referred to as an almost Cauchy-Riemann (CR) structure. If an almost CR structure satisfies the integrability condition
\begin{equation}\label{eq:THF_integrability}
  [T^{1,0} \oplus \bC v, T^{1,0} \oplus \bC v] \subseteq T^{1,0} \oplus \bC v,
\end{equation}
one speaks of a THF.\footnote{The similar integrability condition $[T^{1,0}, T^{1,0}] \subseteq T^{1,0}$ defines a CR manifold.} Intuitively, $m_{ij}$ determines how $\Phi$ is imbedded in $\Theta^{ij}$ and thus how $T^{1,0}$ is embedded in $TM_H$. If one forgets about the vertical direction $v$ for a moment, the question of integrability of $\Phi$ is similar to the question under which circumstances a quaternion K\"ahler structure on a four-manifold admits an integrable complex structure.

To address the question of the existence of an $m_{ij}$ satisfying \eqref{eq:THF_integrability} we follow the construction of \cite{Pan:2015nba} and define the projection operator
\begin{equation}
  H^i_{\phantom{i}j} = (\det m)^{-1/2} m^i_{\phantom{i}j} - \imath \delta^i_{\phantom{i}j}.
\end{equation}
One can then show that
\begin{equation}\label{eq:spinorial_holomorphy_condition}
  X \in T^{1,0} \oplus \bC v \qquad \Leftrightarrow \qquad
  X^\mu H^i_{\phantom{i}j} \Pi_\mu^\nu \gamma_\nu \epsilon^j = 0,
\end{equation}
if the supersymmetry spinor $\epsilon^i$ satisfies a reality condition such as \eqref{eq:symplectic_Majorana}.
Acting from the left with $\mathcal{D}_Y$ for $Y \in T^{1,0} \oplus \bC v$ and antisymmetrizing over $X, Y$, one derives the spinorial integrability condition
\begin{equation}\label{eq:spinorial_integrability_condition}
  [X, Y] \in T^{1,0} \oplus \bC v \qquad \Leftrightarrow \qquad
  X^{[\mu} Y^{\nu]} \mathcal{D}_\mu (H^i_{\phantom{i}j} \Pi_\nu^\rho \gamma_\rho \epsilon^j) = 0.
\end{equation}
Note that $H^i_{\phantom{i}j}$ satisfies $H^2 = -2\imath H$ and has eigenvalues $0$ and $-2\imath$. Thus, $H^i_{\phantom{i}j} \epsilon^j$ projects the doublet $\epsilon^i$ to a single spinor that is a linear combination of the two. It is this spinor that will define the THF.

To proceed, we first consider $X, Y \in T^{1,0}$. After substituting \eqref{eq:intrinsic_torsions_introduced} and making repeatedly use of \eqref{eq:spinorial_holomorphy_condition}, one finds that the condition \eqref{eq:spinorial_integrability_condition} reduces to the vanishing of
\begin{equation}
  X^{[\mu} Y^{\nu]} (\partial_\mu H^i_{\phantom{i}j} + [Q_\mu, H]^i_{\phantom{i}j}) \gamma_n \epsilon^j.
\end{equation}
Similarly, the case $X \in T^{1,0}$, $Y = v$ leads to the condition that
\begin{equation}
  X^m (\partial_v H^i_{\phantom{i}j} + [\iota_v Q, H]^i_{\phantom{i}j}) \gamma_m \epsilon^j
\end{equation}
must be identically zero. Contracting both expressions with $\epsilon^j$ and symmetrizing over $\SU(2)$ indices, we conclude that the integrability condition \eqref{eq:THF_integrability} can only be satisfied if and only if
\begin{equation}\label{eq:THF_integrability_condition_for_connection}
  \mathcal{D}^Q_\mu H^i_{\phantom{i}j} \equiv \partial_\mu H^i_{\phantom{i}j} + [Q_\mu, H]^i_{\phantom{i}j} = 0,
\end{equation}
i.e.~iff the projection $H^i_{\phantom{i}j}$ is covariantly constant with respect to the connection defined by $Q$.

From the condition that $H^i_{\phantom{i}j}$ be covariantly constant we derive the necessary condition that it is also annihilated by the action of the corresponding curvature tensor:
\begin{equation}\label{THF_integrability_condition_for_curvature}
  [R^Q_{\mu\nu}, H]^i_{\phantom{i}j} = 0,
\end{equation}
where $R^Q_{\mu\nu} = [\mathcal{D}^Q_\mu, \mathcal{D}^Q_\nu]$. Now, one can only solve \eqref{THF_integrability_condition_for_curvature} if the $\SU(2)$ curvature $R^Q$ lies in a $\U(1)$ inside $\SU(2)_R$. Note that since the curvature $R^Q$ arises from the intrinsic torsions, we can relate it to the Riemann tensor and $P_{\mu\nu}$ using \eqref{eq:intrinsic_torsions_introduced}. The resulting expression is not too illuminating however.

To conclude we will relate the integrability condition \eqref{eq:THF_integrability_condition_for_connection} to the findings of \cite{Pan:2015nba}. There it was found that solutions of the $\mathcal{N}=1$ Poincar\'e supergravity of \cite{Kugo:2000hn,Kugo:2000af,Zucker:1999ej} define THFs if $m_{ij} = t_{ij}$ and $\forall X \in TM_H, \mathcal{D}_X t_{ij} = 0$. In other words, the unique choice for $m_{ij}$ is the field $t_{ij}$ appearing in the Weyl multiplet of that theory and the latter has to be covariantly constant (with respect to the usual $\SU(2)_R$ connection $V_\mu^{ij}$) along the horizontal leaves of the foliation. To relate our results to this, consider the case where $v$ is actually Killing. It follows that we can assume $\pounds_v H^i_{\phantom{i}j} = 0$ and thus the vertical part of \eqref{eq:THF_integrability_condition_for_connection} takes the form of the first condition of \cite{Pan:2015nba}, namely
\begin{equation}
  [\iota_v Q, H]^i_{\phantom{i}j} = 0.
\end{equation}
Moreover, our general solutions \eqref{eq:dilatino_solution_Delta_ODE} and \eqref{eq:W} are solved by $W = 0$; while this solution is not unique, it makes the connection to \cite{Pan:2015nba} very eveident as it follows now that $\Pi(Q)^{ij} = \Pi(V)^{ij}$ and so the horizontal part of \eqref{eq:THF_integrability_condition_for_connection} reproduces the second condition from \cite{Pan:2015nba}:
\begin{equation}
  \forall X \in T^{1,0} \qquad \mathcal{D}_X H^i_{\phantom{i}j} = X^\mu (\partial_\mu H^i_{\phantom{i}j} + [V_\mu, H]^i_{\phantom{i}j}) = 0.
\end{equation}

\section{Examples} \label{sec:examples}

Let us now discuss some specific examples illustrating the general results from the previous sections.

\subsection{Flat $\mathbb{R}^5$}\label{sec:example_flat_R5}

Flat space admits constant spinors generating the Poincar\'e supersymmetries. In addition, we can consider the spinor generating superconformal supersymmetries $\epsilon^i = x_\mu \gamma^\mu \epsilon_0^i$, where $\epsilon_0^i$ is constant. Let us see how these fit into our general set-up. For the Poincar\'e supersymmetries, it is clear that we just have $Q=P=V=T=0$. For the superconformal spinors on the other hand, the gravitino and dilatino equations are solved by
\begin{equation}
  \eta^i = - \imath \epsilon_0^i, \qquad
  T_{\mu\nu} = V_\mu^{ij} = D = 0.
\end{equation}
The intrinsic torsions are
\begin{equation}
  Q_\mu^{ij} = - \frac{2}{s x^2} x_\kappa \Theta^{ij\kappa}_{\phantom{ij\kappa}\mu}, \qquad
  P_{[\mu\nu]} = \frac{1}{s x^2} (x \wedge v)_{\mu\nu}, \qquad
  P_{(\mu\nu)} = s^{-1} x_\kappa v^\kappa \delta_{\mu\nu}.
\end{equation}
Note that $\Theta_{\nu\rho}^{ij} Q^\rho_{ij} = - \frac{3s}{x^2} \Pi_\nu^\sigma x_\sigma$ and thus
\begin{equation}
  \frac{2}{3} 2 v_{[\mu} \Theta^{ij}_{\nu]\rho} Q^\rho_{ij} = \frac{s}{x^2} (\Pi_\mu^\rho v_\nu - \Pi_\nu^\rho v_\mu) x_\rho = \frac{s}{x^2} (x\wedge v)_{\mu\nu} = s^2 P_{[\mu\nu]}.
\end{equation}
We don't only see that \eqref{eq:gravitino_solution_i} is satisfied, yet also that the only contribution to the right hand side of that equation comes from $\frac{2}{3} 2 v_{[\mu} \Theta^{ij}_{\nu]\rho} Q^\rho_{ij}$ while it is exactly the term that vanishes, $\Theta^{ij}_{\mu\nu} v_\rho Q^\rho_{ij}$, that contributes in the in the Sasaki-Einstein case to be discussed below.

Note that the superconformal supersymmetries involve non-zero trace of $P$. Hence, these supersymmetries are broken by the background scalar VEV corresponding to $g_{YM}^{-2}$. This just reflects the general wisdom that the 5d YM coupling, being dimensionful, breaks conformal invariance.

\subsection{$\mathbb{R}\times S^4$}

Consider now $\mathbb{R}\times S^4$, with $\mathbb{R}$ parametrized by $x^5=\tau$ and $v$ not along $\frac{\partial}{\partial\tau}$. As described in \cite{Kim:2012gu} -- where the explicit spinor solutions are written as well, the spinors satisfy
\begin{equation}
\nabla_{\mu}\epsilon^q=-\frac{1}{2}\gamma_{\mu}\,\gamma_5\epsilon^q\, ,\qquad \nabla_{\mu}\epsilon^s=\frac{1}{2}\gamma_{\mu}\,\gamma_5\epsilon^s.
\end{equation}
Here $\epsilon^{q,\,s}$ generate Poincar\'e and superconformal supersymmetries respectively. It is straightforward to see that these solutions fit in our general scheme with
\begin{equation}
Q^{ij}_{\mu}= \pm\frac{1}{2s} w_{\kappa}\Theta^{ij\kappa}_{\phantom{ij\kappa}\mu},\qquad P_{[\mu\nu]}=\mp \frac{1}{2s}\,(w\wedge v)_{\mu\nu}, \qquad P_{(\mu\nu)}=\mp \frac{1}{2s}\,w_{\kappa} v^{\kappa}\,g_{\mu\nu},
\end{equation}
where upper signs correspond to the $\epsilon_q$ while lower signs correspond to the $\epsilon_s$. In addition we have defined $w=d\tau$. Note that the trace of $P$ does not vanish, implying that $v$ is conformal Killing. Thus this is a genuine solution of superconformal supergravity that cannot be embedded in $\mathcal{N}=1$ Poincar\'e supergravity. Moreover, as discussed above, this implies that no (constant) Yang-Mills coupling can be turned on on this background (see \cite{Kim:2014kta} for a further discussion in the maximally supersymmetric case).

\subsection{Topological twist on $\mathbb{R}\times M_4$}

Manifolds of the form $\mathbb{R}\times M_4$ can be regarded as supersymmetric backgrounds at the expense of turning on a non-zero $V$ such that the spinors are gauge-covariantly constant

\begin{equation}
\mathcal{D}_{\mu}\epsilon^i=0.
\end{equation}
To show that we consider $v=\partial_{\tau}$, being $\tau$ the coordinate parametrizing $\mathbb{R}$. Then, from \eqref{eq:intrinsic_torsions_in_terms_of_covariant_derivative}, it follows that $P_{\mu\nu}=0$. Furthermore, by choosing $V=Q$ -- which translates into $W_{\mu}=0$, $\Delta^{ij}=0$ and implies $T_{\mu\nu}=0$ -- all the remaining constraints are automatically solved. This is nothing but the topological twist discussed in \cite{Rodriguez-Gomez:2015xwa} (see also \cite{Anderson:2012ck} for the maximally supersymmetric case; twisted theories on five manifolds were also considered in \cite{Bak:2015hba}). Note that since $P=0$, in these backgrounds the Yang-Mills coupling can indeed be turned on.

\subsection{$\SU(2)_R$ twist on $M_5$}

If $M_5$ is not a direct product, one can still perform an $SU(2)_R$ twist. For $v$ Killing, the details of this can be found in \cite{Pan:2015nba}. One can perform an identical calculation for the conformal supergravity in question. In the case of a $\mathbb{R}$ or $\U(1)$ bundle over some $M_4$ for example, one finds $T$ to be the curvature of fibration.

\subsection{Sasaki-Einstein manifolds}

For a generic Sasaki-Einstein manifold the spinor satisfies
\begin{equation}
  \nabla_\mu \epsilon_i = -\frac{\imath}{2} \gamma_\mu (\sigma^3)_i^{\phantom{i}j} \epsilon_j.
\end{equation}
It follows that
\begin{equation}
  P_{\mu\nu} = -\frac{\imath}{2} s^{-1} (\sigma^3_{ij} \Theta^{ij})_{\mu\nu}, \qquad
  Q_\mu^{ij} = - \frac{\imath}{2} s^{-1} v_\mu (\sigma^3)^{ij}.
\end{equation}
Clearly
\begin{equation}
  s^2 P_{[\mu\nu]} = -\frac{\imath}{2} s (\sigma^3_{ij} \Theta^{ij})_{\mu\nu} = \Theta_{\mu\nu}^{ij} Q^\rho_{ij} v_\rho.
\end{equation}
Hence, upon taking $V_\mu^{ij} = 0 = T_{\mu\nu}$, we indeed have a solution of \eqref{eq:gravitino_solution_i} and \eqref{eq:gravitino_solution_ii}. 

Note that the trace of $P$ is vanishing, and hence in these backgrounds the Yang-Mills coupling can be turned on. This holds also for Sasakian manifolds. Super Yang-Mills theories on these were considered in e.g.~\cite{Qiu:2013pta}.

\subsection{$S^5$}

The $S^5$ case is paticularly interesting as well, as it leads to the supersymmetric partition function \cite{Hosomichi:2012ek,Kallen:2012va}. Not surprisingly, since $S^5$ can be conformally mapped into $\mathbb{R}^5$, the solution fits into our general discussion including two sets of spinors, one corresponding to the Poincar\'e supercharges and the other corresponding to the superconformal supercharges. Writing the $S^5$ metric as that of conformally $S^5$ as
\begin{equation}
ds^2=\frac{4}{(1+\vec{x}^2)^2}\,d\vec{x}^2, 
\end{equation}
we find for the Poincare supersymmetries
\begin{equation}
Q^{ij}_{\mu}= \frac{1}{2s} x_{\kappa}\Theta^{ij\kappa}_{\phantom{ij\kappa}\mu},\qquad P_{[\mu\nu]}=- \frac{1}{2s}\,(x\wedge v)_{\mu\nu}, \qquad P_{(\mu\nu)}=- \frac{1}{2s}\,x_{\kappa} v^{\kappa}\,g_{\mu\nu}.
\end{equation}
For the superconformal supercharges on the other hand, we find
\begin{equation}
Q^{ij}_{\mu}= -\frac{1}{2sx^2} x_{\kappa}\Theta^{ij\kappa}_{\phantom{ij\kappa}\mu},\qquad P_{[\mu\nu]}=\frac{1}{2sx^2}\,(x\wedge v)_{\mu\nu}, \qquad P_{(\mu\nu)}= \frac{1}{2sx^2}\,x_{\kappa} v^{\kappa}\,g_{\mu\nu}.
\end{equation}
Note that in both these cases the trace of $P$ is non-zero, so neither of these spinors are preserved if we deform the theory with a Yang-Mills coupling. Nevertheless it is possible to find a combination of supercharges which does allow for that. This can be easily understood by looking at the explicit form of the spinors, which in these coordinates is simply
\begin{equation}
\epsilon_q^i=\frac{1}{\sqrt{1+\vec{x}^2}}\epsilon_0^i,\qquad \epsilon_s^i=\frac{1}{\sqrt{1+\vec{x}^2}}\slashed{x}\eta_0^i,
\end{equation}
being $\epsilon_0^i$ and $\eta_0^i$ constant spinors. Considering for instance $\slashed{\nabla}\epsilon_{q}^i\supset P_{[\mu\nu]}\gamma^{\mu}\gamma^{\nu}\epsilon_q^i+P^{\mu}_{\phantom{\mu}\mu}\epsilon^i_q$, we see that 
the term with $P_{[\mu\nu]}$ involves a contraction $\slashed{x}\epsilon_q^i$ which is basically $\epsilon_s^i$. This suggests that one might consider a certain combination of $\epsilon_q$ and $\epsilon_s$ for which the effective $P$-trace is a combination of $P_{[\mu\nu]}$ and $P^\mu_{\phantom{\mu}\mu}$ which might vanish. Indeed one can check that this is the case. Choosing for instance the Majorana doublet $\xi^i$ constructed as
\begin{equation}
\xi^1=\epsilon_q^1+\epsilon_s^2,\qquad \xi^2=\epsilon_q^2-\epsilon_s^1,
\end{equation}
it is easy to see that it satisfies 
\begin{equation}
  \nabla_\mu \epsilon_i = -\frac{\imath}{2} \gamma_\mu (\sigma^2)_i^{\phantom{i}j} \epsilon_j;
\end{equation}
that is, the same equation as that for the Sasaki-Einstein case. Therefore, borrowing our discussion above, it is clear that it admitts a Yang-Mills kinetic term. Indeed, this is corresponds, up to conventions, to the choice made in \cite{Hosomichi:2012ek,Kallen:2012va} to compute the supersymmetric partition function.

\section{Conclusions}\label{sec:conclusions}

In this paper we have studied general solutions to $\mathcal{N}=2$ conformal supergravity. In the spirit of \cite{Festuccia:2011ws}, these provide backgrounds admitting five-dimensional supersymmetric quantum field theories. The starting point of our analysis, being conformal supergravity, requires that such quantum field theories must exhibit conformal invariance. In particular, the action for vector multiplets must be the cubic completion of 5d Chern-Simons term instead of the standard quadratic Maxwell one. However, since the Yang-Mills coupling can be thought as a VEV for the scalar in a background vector multiplet, we can regard gauge theories as conformal theories conformally coupled to background vector multiplets whose VEVs spontaneously break conformal invariance. From this perspective it is very natural to consider superconformal supergravity as the starting point to construct the desired supersymmetric backgrounds.

We have described the most generic solution to $\mathcal{N}=2$ five-dimensional conformal supergravity (see also \cite{Kuzenko:2014eqa}). By expanding spinor covariant derivatives in intrinsic torsions we have been able to find a set of algebraic equations \eqref{eq:symmetric_part_of_P}, \eqref{eq:Q-V_parametrization}, \eqref{eq:T-solution}, \eqref{eq:D} together with a set of differential constraints \eqref{eq:dilatino_solution_Delta_ODE}, \eqref{eq:W} characterizing the most general solution. Interestingly, the solutions admit transverse holomorphic foliations if the $\SU(2)_R$ connection $R^Q$ ``abelianizes'' by lying along a $\U(1)$ inside $\SU(2)_R$, in agreement with the discussion in \cite{Alday:2015lta}.

On general grounds, the only obstruction to the existence of supersymmetric backgrounds is the requirement of a conformal Killing vector. On the other hand we have showed that only when the vector becomes actually Killing a constant VEV for background vector multiplet scalars can be turned on. This shows that all cases where a Yang-Mills theory with standard Maxwell kinetic term can be supersymmetrically constructed are in fact captured by Poincar\'e supergravity. On the other hand, on backgrounds admitting only a conformal Killing vector we can still turn on a Yang-Mills coupling at the expense of being position-dependent. While this is certainly non-standard, in particular this allows to think of the quadratic part of the Yang-Mills action as the regulator in index computations.

Having constructed all supersymmetric backgrounds of $\mathcal{N}=2$ superconformal supergravity, the natural next step would be the computation of supersymmetric partition functions. In particular, it is natural to study on what data they would depend along the lines of \textit{e.g.} \cite{Closset:2013vra}. For initial progress in this direction see \cite{Imamura:2014ima,Pan:2015nba}. We postpone such study for future work.

\section*{Acknowledgements}
The authors would like to thank S.~Kuzenko, P.~Meessen, Y.~Pan and D.~Rosa for useful conversations and correspondence. The authors are partly supported by the spanish grant MINECO-13-FPA2012-35043-C02-02. In addition, they acknowledge financial support from the Ramon y Cajal grant RYC-2011-07593 as well as the EU CIG grant UE-14-GT5LD2013-618459. The work of A.P~is funded by the Asturian government's SEVERO OCHOA grant BP14-003. The work of J.S.~is funded by the Asturian government's CLARIN grant ACB14-27. A.P. would like to acknowledge the String Theory Group of the Queen Mary University of London and  the String Theory Group of the Imperial College of London for very kind hospitality during the final part of this project. Moreover the authors would like to acknowledge the support of the COST Action MP1210 STSM.

\appendix

\section{Conventions}\label{sec:conventions}

We use the standard NE-SW conventions for $\SU(2)_R$ indices $\{ i, j, k, l \}$ with $\epsilon^{12} = \epsilon_{12} = 1$. The charge conjugation matrix $C$ is antisymmetric, hermitian and orthogonal, i.e.~$C^* = C^T = -C = C^{-1}$. Its action on gamma matrices is given by $(\gamma^a)^* = (\gamma^a)^T = C \gamma^a C^{-1}$. In general we choose not to write the charge conjugation matrix explicitly; thus $\epsilon^i \eta^j = (\epsilon^i)^T C \eta^j$. Antisymmetrised products of gamma matrices are defined with weight one,
\begin{equation}
  \gamma_{a_1 \dots a_p} = \frac{1}{p!} \gamma_{[a_1} \dots \gamma_{a_p]},
\end{equation}
yet contractions between tensors and gamma matrices are not weighted.
\begin{equation}
  \gamma \cdot T = \gamma^{\mu\nu} T_{\mu\nu}.
\end{equation}
In general, symmetrization $T_{(\mu_1 \dots \mu_p)}$ and antisymmetrization $T_{[\mu_1 \dots \mu_p]}$ are with weight one however.

One can impose a symplectic Majorana condition
\begin{equation}\label{eq:symplectic_Majorana}
  \epsilon^{ij} (\epsilon^j)^* = C \epsilon^i,
\end{equation}
yet as we mentioned in the main body of this paper it is generally sufficient for us to assume $s$ to be non-vanishing and $v$ to be real.

Using Fierz identities, one finds the following identities involving the spinor bilinear $\Theta^{ij}$:
\begin{IEEEeqnarray}{rCl}
  \Theta^{ij}_{\mu\nu} \Theta^{kl\mu\nu} &=& s^2 (\epsilon^{ik} \epsilon^{jl} + \epsilon^{il} \epsilon^{jk}), \\
  \Theta_{\kappa\lambda}^{ij} \Theta^{\mu\nu}_{ij} &=& \frac{s^2}{2} (\Pi_\kappa^\mu \Pi_\lambda^\nu - \Pi_\kappa^\nu \Pi_\lambda^\mu) - \frac{s}{2} \epsilon_{\kappa\lambda}^{\phantom{\kappa\lambda}\mu\nu\rho} v_\rho \label{eq:double_theta_identity}\\
  \Theta_{\mu\rho}^{ij} \Theta^{kl \rho\nu} &=& -\frac{s^2}{4} (\epsilon^{ik} \epsilon^{jl} + \epsilon^{il} \epsilon^{jk}) \Pi_\mu^{\phantom{\mu}\nu} + \frac{s}{4} (\epsilon^{jk} \Theta^{il} + \epsilon^{ik} \Theta^{jl} + \epsilon^{jl} \Theta^{ik} + \epsilon^{il} \Theta^{jk})_\mu^{\phantom{\mu}\nu} \label{eq:double_theta_identity_2}.
\end{IEEEeqnarray}

\section{Details of the computation}\label{sec:computation}

In this appendix we summarize the most relevant details of the computation that lead us to the two equations \eqref{eq:gravitino_solution_i} and \eqref{eq:gravitino_solution_ii} and to the three differential equations \eqref{eq:dilatino_solution_Delta_ODE}, \eqref{eq:D} and \eqref{eq:W}, that allow to determine $\Delta^{ij}$, the scalar $D$ and the vector $W_{\kappa}$.

\subsection{Gravitino equation}
In this subsection we furnish further details for the derivation of the equations  \eqref{eq:gravitino_solution_i} and \eqref{eq:gravitino_solution_ii}. As explained in section \ref{sec:general_solutions} we rewrite the covariant derivative acting on the spinor $\epsilon^{i}$ as
\begin{equation}
\label{eq:cov}
\mathcal{D}_\mu \epsilon^i = \nabla_\mu \epsilon^i - V^{ij}_{\mu}\epsilon_j = P_{\mu\nu} \gamma^{\nu}\epsilon^i + (Q-V)_\mu^{ij}\epsilon_j.
\end{equation}
Inserting this expression for the covariant derivative in the gravitino equation \eqref{eq:gravitino-variation_w/out_eta} we obtain
\begin{IEEEeqnarray}{rCl}
\label{eq:gravitino_cov}
  0 &=& \frac{3}{4} P_{[\mu\nu]} \gamma^\nu \epsilon^i + \frac{1}{8} \epsilon_{\mu\kappa\lambda\sigma\tau} P^{[\kappa\lambda]} \gamma^{\sigma\tau} \epsilon^i + \frac{5}{4} P_{(\mu\nu)} \gamma^\nu \epsilon^i - \frac{1}{4} \gamma_\mu P^\nu_{\phantom{\nu}\nu} \epsilon^i \nonumber\\
  &&+ (Q_\mu^{ij} - V_\mu^{ij}) \epsilon_j - \frac{1}{4} \gamma_{\mu\nu} (Q^{\nu ij} - V^{\nu ij}) \epsilon_j 
  - \frac{\imath}{2} \epsilon_{\mu\kappa\lambda\sigma\tau} T^{\kappa\lambda} \gamma^{\sigma\tau} \epsilon^i - 3 \imath T_{\mu\nu} \gamma^\nu \epsilon^i \nonumber\\
  &=& \frac{5}{4} \left( P_{(\mu\nu)} - \frac{1}{5} g_{\mu\nu} P^\lambda_{\phantom{\lambda}\lambda} \right) \gamma^\nu \epsilon^i + (Q-V)_\mu^{ij} \epsilon_j - \frac{1}{4} \gamma_{\mu\nu} (Q - V)^{\nu ij} \epsilon_j \nonumber\\
  &&+ \frac{3}{4} (P_{[\mu\nu]} -4\imath T_{\mu\nu}) \gamma^\nu \epsilon^i + \frac{1}{8} \epsilon_{\mu\kappa\lambda\sigma\tau} (P^{[\kappa\lambda]} - 4\imath T^{\kappa\lambda}) \gamma^{\sigma\tau} \epsilon^i.
\label{eq:gravitino_variation_in_terms_of_intrinsic_torsions}
\end{IEEEeqnarray}
We manipulate the previous expression, as discussed in section \ref{sec:general_solutions}, multiplying it from the left by $\epsilon_i \gamma_\kappa$. In this way we obtain the equation \eqref{eq:gravitino_via_torsion_vector_contraction}. While  we obtain the equation \eqref{eq:gravitino_via_torsion_triplet_contraction} multiplying the equation \eqref{eq:gravitino_cov} by $\epsilon^j$ and symmetrizing in the indices $i$ and $j$.

In order to recover the equation \eqref{eq:gravitino_solution_i} we have to determine $(P - 4 \imath T)^+$ and $(P - 4 \imath T)^-$. Therefore  we project the equation \eqref{eq:gravitino_via_torsion_vector_contraction} on the horizontal space using the projector operator $\Pi^{\mu}_{\nu} = \delta^{\mu}_{\nu} -s^{2}v^{\mu}v_{\nu}$. We find
\begin{equation}
 0 = \frac{5}{8} \Pi_\mu^\kappa \Pi_\nu^\lambda \left[ (P - 4 \imath T)_{[\kappa\lambda]} + \frac{1}{2} \epsilon_{\kappa\lambda\sigma\tau\rho} (P - 4 \imath T)^{\sigma\tau} v^\rho \right] = \frac{5}{4} (P - 4 \imath T)^+.
\end{equation}
This means that $\Pi(P-4iT)$ is anti-self dual. On the other hand contracting the equation \eqref{eq:QVPT_identity_triplet} with $\Theta_{ij\kappa\lambda}$ and using the identity \eqref{eq:double_theta_identity} we get 
\begin{equation}
  0 = s^3 \left( \Pi_{\kappa\mu} \Pi_{\lambda\nu} - \frac{1}{2} s^{-1} \epsilon_{\kappa\lambda\mu\nu\rho} v^\rho \right) (P - 4\imath T)^{[\kappa\lambda]} - 2 s \Theta_{ij\kappa\lambda} (Q -V)^{ij}_\rho v^\rho.
\end{equation}
Solving the previous expression we obtain $(P - 4\imath T)^- = s^{-2} \Theta^{ij} \imath_v (Q - V)_{ij}$. Therefore we know all the components of $(P-4iT)$, since  we have an equation for $(P-4iT)^{+}$, an equation for $(P-4iT)^{-}$ and finally an equation for $\imath_v(P-4\imath T)$. Putting these information together we recover the equation \eqref{eq:gravitino_solution_i}.

In order to determine the equation \eqref{eq:gravitino_solution_ii} we evaluate the  projection of the equation \eqref{eq:gravitino_via_torsion_triplet_contraction}, using the identities \eqref{eq:QVPT_identity_vector} and \eqref{eq:QVPT_identity_triplet} we get
\begin{equation}
  0 = \frac{1}{2} s \Pi_\mu^\nu (Q - V)_\nu^{ij} + \frac{1}{4} (Q - V)^{\nu (j}_{\phantom{\nu (j}k} \Theta^{i)k}_{\mu\nu} + \frac{1}{6} s^{-1} \Theta_{\mu\nu}^{ij} \Theta^{\nu\rho}_{kl} (Q -V)^{kl}_\rho.
\end{equation}
Using the identity \eqref{eq:double_theta_identity_2} the previous expression becomes
\begin{equation}
  0 = s \Pi_\mu^{\phantom{\mu}\nu} (Q-V)_{\nu\phantom{i}j}^{\phantom{\nu}i} + \frac{1}{2} [(Q-V)^\nu, \Theta_{\mu\nu}]^i_{\phantom{i}j}.
\end{equation}
Obtaining in this way the equation \eqref{eq:gravitino_solution_ii}.

\subsection{Dilatino equation}
In this subsection we furnish further details regarding the derivation of the equations \eqref{eq:dilatino_solution_Delta_ODE}, \eqref{eq:D} and \eqref{eq:W}. The most involved terms that appear in the equation \eqref{eq:dilatino-variation_D-squared-form} are
\begin{IEEEeqnarray}{rCl}
  \mathcal{D}^\mu \mathcal{D}_\mu \epsilon^i &=& \frac{1}{5} \slashed{\nabla} P^\mu_{\phantom{\mu}\mu} \epsilon^i - \gamma^\mu \nabla^\nu P_{[\mu\nu]} \epsilon^i + \frac{1}{5} (P^\mu_{\phantom{\mu}\mu})^2 \epsilon^i + P_{[\mu\nu]} P^{[\mu\nu]} \epsilon^i \nonumber\\
  &&- \nabla^\mu (Q-V)_{\mu\phantom{i}j}^{\phantom{\mu}i} \epsilon^j - V_{\mu\phantom{i}j}^{\phantom{\mu}i} (Q-V)_{\phantom{\mu j}k}^{\mu j} \epsilon^k + (Q-V)_{\mu\phantom{i}j}^{\phantom{\mu}i} Q_{\phantom{\mu j}k}^{\mu j} \epsilon^k \nonumber\\
  &&+ \frac{2}{5} P^\kappa_{\phantom{\kappa}\kappa} \gamma^\mu (Q-V)_\mu^{ij} \epsilon_j - 2 \gamma^\mu P_{[\mu\nu]} (Q-V)^{\nu ij} \epsilon_j
\end{IEEEeqnarray}
and
\begin{IEEEeqnarray}{rCl}
  \gamma_{\kappa\lambda\mu} T^{\kappa\lambda} \mathcal{D}^\mu \epsilon^i &=& \gamma_{\kappa\lambda\mu\nu} T^{\kappa\lambda} P^{[\mu\nu]} \epsilon^i + \frac{3}{5} P^\kappa_{\phantom{\kappa}\kappa} T_{\mu\nu} \gamma^{\mu\nu} \epsilon^i - 2 P_{[\mu\kappa]} T^\kappa_{\phantom{\kappa}\nu} \gamma^{\mu\nu} \epsilon^i \nonumber\\
  &&+ \gamma_{\kappa\lambda\mu} T^{\kappa\lambda} (Q - V)^{\mu ij} \epsilon_j,\\
  \gamma^\mu T_{\mu\nu} \mathcal{D}^\nu \epsilon^i &=& - P_{[\mu\nu]} T^{\mu\nu} \epsilon^i - P_{[\mu\kappa]} T^\kappa_{\phantom{\kappa}\nu} \gamma^{\mu\nu} \epsilon^i + \frac{1}{5} P^\kappa_{\phantom{\kappa}\kappa} T_{\mu\nu} \gamma^{\mu\nu} \epsilon^i + \gamma^\mu T_{\mu\nu} (Q-V)^{\nu ij} \epsilon_j.
\end{IEEEeqnarray}
 
\paragraph{The symmetric contraction}

Multiplying the equation \eqref{eq:dilatino-variation_D-squared-form} by $\epsilon^{j}$ and symmetrizing in $i$ and $j$ we obtain
\begin{IEEEeqnarray}{rCl}
  0 &=& \frac{1}{8} \epsilon^{(i} \mathcal{D}^\mu \mathcal{D}_\mu \epsilon^{j)} + \frac{3\imath}{40} \epsilon^{(i} \gamma_{\kappa\lambda\mu} T^{\kappa\lambda} \mathcal{D}^\mu \epsilon^{j)} + \frac{11\imath}{40} \epsilon^{(i} \gamma^\mu T_{\mu\nu} \mathcal{D}^\nu \epsilon^{j)} \nonumber\\
  &&+ \frac{\imath}{4} \epsilon^{(i} \gamma_{\kappa\lambda\mu} \epsilon^{j)} \nabla^\mu T^{\kappa\lambda}.
\end{IEEEeqnarray}
The individual components are
\begin{IEEEeqnarray}{rCl}
  \epsilon^{(i} \mathcal{D}^\mu \mathcal{D}_\mu \epsilon^{j)} &=& 
  \frac{s}{2} \left[ \nabla^\mu (Q-V)_\mu^{ij} + (Q+V)_{\mu\phantom{(i}k}^{\phantom{\mu}(i} (Q-V)^{\mu j) k} \right] \nonumber\\
  &&+ \frac{1}{5} P^\kappa_{\phantom{\kappa}\kappa} v_\mu (Q-V)^{\mu ij} - v^\mu P_{[\mu\nu]} (Q-V)^{\nu ij}, \\
  \epsilon^{(i} \gamma_{\kappa\lambda\mu} T^{\kappa\lambda} \mathcal{D}^\mu \epsilon^{j)} &=& \frac{3}{5} P^\kappa_{\phantom{\kappa}\kappa} T^{\mu\nu} \Theta^{ij}_{\mu\nu} - 2 P_{[\mu\kappa]} T^\kappa_{\phantom{\kappa}\nu} \Theta^{ij\mu\nu} + \frac{1}{2} \epsilon_{\kappa\lambda\mu}^{\phantom{\kappa\lambda\mu}\nu\rho} T^{\kappa\lambda} \Theta^{k(i}_{\nu\rho} (Q-V)^{\mu j)}_{\phantom{\mu j)}k}, \\
  \epsilon^{(i} \gamma^\mu T_{\mu\nu} \mathcal{D}^\nu \epsilon^{j)} &=& -P_{[\mu\kappa]} T^\kappa_{\phantom{\kappa}\nu} \Theta^{ij\mu\nu} + \frac{1}{5} P^\mu_{\phantom{\mu}\mu} T^{\kappa\lambda} \Theta^{ij}_{\kappa\lambda} + \frac{1}{2} v_\mu T^{\mu\nu} (Q-V)^{ij}_\nu,\\
  \epsilon^{(i} \gamma_{\kappa\lambda\mu} \epsilon^{j)} \nabla^\mu T^{\kappa\lambda} &=& -\frac{1}{2} \epsilon_{\kappa\lambda\mu}^{\phantom{\kappa\lambda\mu}\nu\rho} \Theta_{\nu\rho}^{ij} \nabla^\mu T^{\kappa\lambda}.
\end{IEEEeqnarray}
Putting the various terms together we recover the expression \eqref{eq:dilatino_solution_Delta_ODE}.

\paragraph{The vector contraction}

Multiplying the equation \eqref{eq:dilatino-variation_D-squared-form} with $\epsilon^i \gamma_\mu$ and contracting we obtain
\begin{IEEEeqnarray}{rCl}
  0 &=& v_\mu \left( \frac{32 D + R}{128} + \frac{1}{15} T_{\mu\nu} T^{\mu\nu} \right) + \frac{1}{8} \epsilon^i \gamma_\mu \mathcal{D}^\nu \mathcal{D}_\nu \epsilon_i + \frac{3\imath}{40} \epsilon^i \gamma_\mu \gamma_{\kappa\lambda\nu} T^{\kappa\lambda} \mathcal{D}^\nu \epsilon_i \nonumber\\
  &&+ \frac{11\imath}{40} \epsilon^i \gamma_\mu \gamma^\kappa T_{\kappa\lambda} \mathcal{D}^\lambda \epsilon_i + \frac{\imath}{4} \epsilon_\mu^{\phantom{\mu}\nu\kappa\lambda\sigma} v_\sigma \nabla_\nu T_{\kappa\lambda} + \frac{\imath s}{2} \nabla^\nu T_{\mu\nu} - \frac{s}{5} \epsilon_\mu^{\phantom{\mu}\kappa\lambda\sigma\tau} T_{\kappa\lambda} T_{\sigma\tau}.
\end{IEEEeqnarray}
The most involved terms are given by
\begin{IEEEeqnarray}{rCl}
  \epsilon^i \gamma_\mu \mathcal{D}^\nu \mathcal{D}_\nu \epsilon_i &=& \frac{s}{5} \nabla_\mu P^\kappa_{\phantom{\kappa}\kappa} - s \nabla^\nu P_{[\mu\nu]} - \frac{2}{5} P^\kappa_{\phantom{\kappa}\kappa} \Theta^{ij}_{\mu\nu} (Q-V)^\nu_{ij} + 2 \Theta^{ij}_{\mu\nu} P^{[\nu\rho]} (Q-V)_{\rho ij} \nonumber\\
  &&+ v_\mu \left[ \frac{1}{5} (P^\kappa_{\phantom{\kappa}\kappa})^2 + P_{[\mu\nu]} P^{[\mu\nu]} - \frac{1}{2} (Q-V)_\nu^{ij} (Q-V)^\nu_{ij} \right], \\
  \epsilon^i \gamma_\mu \gamma_{\kappa\lambda\nu} T^{\kappa\lambda} \mathcal{D}^\nu \epsilon_i &=& s \epsilon_{\mu\kappa\lambda\sigma\tau} T^{\kappa\lambda} P^{[\sigma\tau]} + \frac{6}{5} P^\kappa_{\phantom{\kappa}\kappa} T_{\mu\nu} v^\nu - 2 (P_{[\mu\rho]} T^\rho_{\phantom{\rho}\nu} - P_{[\nu\rho]} T^\rho_{\phantom{\rho}\mu}) v^\nu \nonumber\\
  &&- (Q-V)_{\mu ij} \Theta_{\kappa\lambda}^{ij} T^{\kappa\lambda} - 2 T_{\mu\nu} \Theta^{ij\nu\rho} (Q-V)_{\rho ij}, \\
  \epsilon^i \gamma_\mu \gamma^\kappa T_{\kappa\lambda} \mathcal{D}^\lambda \epsilon_i &=& - v_\mu P_{[\kappa\lambda]} T^{\kappa\lambda} - (P_{[\mu\rho]} T^{\rho}_{\phantom{\rho}\nu} - P_{[\nu\rho]} T^{\rho}_{\phantom{\rho}\mu}) v^\nu + \frac{2}{5} P^\kappa_{\phantom{\kappa}\kappa} T_{\mu\nu} v^\nu \nonumber\\
  &&- \Theta^{ij}_{\mu\kappa} T^{\kappa\lambda} (Q-V)_{\lambda ij}.
\end{IEEEeqnarray}
Finally putting the various terms together and projecting on the vertical component we recover the equation \eqref{eq:D}. While projecting on the horizontal component we recover the equation \eqref{eq:W}.

\bibliographystyle{ytphys}
\small\baselineskip=.97\baselineskip
\bibliography{ref}

\end{document}